
\magnification=1200
\baselineskip=6mm
\nopagenumbers
\noindent
\null\vskip 1truecm
\centerline{\bf THE PROJECTIVE UNITARY IRREDUCIBLE REPRESENTATIONS}
\centerline{\bf OF THE POINCAR\'E GROUP IN 1+2 DIMENSIONS}
\vskip 2truecm
\centerline{D.R. Grigore\footnote*{e-mail: grigore@roifa.bitnet}}
\centerline{Department of Theoretical Physics, Institute of
Atomic Physics}
\centerline{Bucharest-Magurele,Romania}
\vskip 2truecm
\centerline{ABSTRACT}
\vskip 1truecm
We give a complete analysis of the projective unitary
irreducible representations of the Poincar\'e group in
1+2 dimensions applying Mackey theorem and using an
explicit formula for the universal covering group
of the Lorentz group in 1+2 dimensions. We provide explicit
formulae for all representations.
\vskip 1truecm
\vfil\eject

\pageno=1
\footline={\hss\tenrm\folio\hss}
{\bf 1. Introduction}

The purpose of this paper is to determine all the projective unitary
irreducible representations of the Poincar\'e group in 1+2
dimensions. The utility of such a study becames evident taking
into account that physics in 1+2 dimensions is nowaday a subject
of active research.

Our analysis is based on Mackey theorem on induced
representations (see e.g. [1]). The computations are in essence
straightforward, but rather tedious because of the
complicated structure of the universal covering group of the
Lorentz group in 1+2 dimensions (i.e. the analogue of
$SL(2,C)$
from 1+3 dimensions). Cohomological arguments ensure that all
the projective representations of the Poincar\'e group in
$1+n$
dimensions
($n > 1$)
are induced by true representations of the corresponding
universal covering group (see e.g. [1],[2]).

To our kowledge, the only attempt to classify all the projective
unitary irreducible representations of the Poincar\'e group in
1+2 dimensions is contained in [3] and is based on a
``sui-generis'' form of Mackey theorem leading to an incomplete
list of representations.

The benefit of a correct application of Mackey theorem is that
it leads to explicit formulae for all the representations we are
looking for.

We divide the analysis in two parts. The main part (mathematical
framework, computation of the orbits, computation of the little
groups, etc.) is concentrated in \$ 2. We defer the analysis of
the projective unitary irreducible representations of the
Lorentz group in 1+2 dimensions to \$ 3. Such an analysis has
been already provided in [4] but explicit formulae are missing
(the interest of the paper is centered on something else and
explicit formulae are not needed but only their existence).
Also we note that the same analysis is done in [5]; however one
considers here the universal covering group of
$SU(1,1) \cong SL(2,R)$.
By comparison, our formulae and proofs seem simpler.
So we think that it is useful to provide a detailed list of
these representations in an explicit way.
In this way, for the benefit of the reader, all the relevant
expressions concerning the projective unitary irreducible
representations of the Poincar\'e group in
$1 + 2$
dimensions will be colected together in a single paper.

The results of this paper can be used to develop the
theory of invariant wave equations on the lines of [6], [7]. We
note that the so-called discrete series of the covering group of
the Lorentz group in 1+2 dimensions can be obtained using
geometric quantization [7].
\vskip 1truecm
{\bf 2. The Poincar\'e group in 1+2 dimensions}

2.1 We denote by $M$ the 1+2-dimensional Minkowski space i.e.
$R^{3}$
with coordinates
$(x^{0},x^{1},x^{2})$
and with the Minkowski bilinear form:
$$
\{x,y\} \equiv x^{0} y^{0} - x^{1} y^{1} - x^{2} y^{2}.\eqno(2.1)
$$

The Lorentz group is:
$$
L \equiv \{ \Lambda \in End(M) \vert \{Lx,Ly\} = \{x,y\}, \forall x,y
\in M \} \eqno(2.2)
$$
considered as a multiplicative group.

We will also consider the orthochronous Lorentz group
$L^{\uparrow} \subset L$:
$$
L^{\uparrow} \equiv \{ \Lambda \in L \vert x^{0} > 0 \Rightarrow
(\Lambda x)^{0} > 0, \forall x \in M \}, \eqno(2.3)
$$
the proper Lorentz group:
$L_{+} \subset L$:
$$
L_{+} \equiv \{ \Lambda \in L \vert det(\Lambda) = 1 \} \eqno(2.4)
$$
and the proper orthochronous Lorentz group:
$$
L^{\uparrow}_{+} \equiv L^{\uparrow} \cap L_{+}. \eqno(2.5)
$$

The Poincar\'e group is a semi-direct product:
$$
P \equiv L \times_{t} M \eqno(2.6)
$$
where we are using the notations of [1]:
$M$
is considered as an additive group and
$t: L \rightarrow Aut(M)$
is simply:
$$
t_{\Lambda} (x) \equiv \Lambda x.
$$

We also have:
$$
P^{\uparrow} \equiv L^{\uparrow} \times_{t} M,
P_{+} \equiv L_{+} \times_{t} M,
P^{\uparrow}_{+} \equiv L^{\uparrow}_{+} \times_{t} M.
$$

2.2 Let us denote. generically, by
$Lie(G)$
the Lie algebra of the Lie group
$G$.
One can prove that
$
H^{2}(Lie(P^{\uparrow}_{+}),R) = 0
$
(see e.g. [2]). Then as in 1+3 dimensions one proves that the
projective representations of
$P^{\uparrow}_{+}$
are induced by true representations of the corresponding
covering group
$\widetilde {P^{\uparrow}_{+}}$
of
$P^{\uparrow}_{+}$.

To construct
$\widetilde{P^{\uparrow}_{+}}$
we need the universal covering group
$\widetilde{L^{\uparrow}_{+}}$
of
$L^{\uparrow}_{+}$.
It is known that
$
SL(2,R) \cong SU(1,1)
$
is a double covering of
$L^{\uparrow}_{+}$.
In [8] one can find an explicit realization for the universal
covering group of
$SU(1,1)$.
{}From this realization one can infer an explicit realization
for the universal covering group of
$SL(2,R)$.
We will prefer to work with
$SL(2,R)$
rather than with
$SU(1,1)$.
As in [9] we define the manifold:
$$
G \equiv R \times D \eqno(2.7)
$$
where:
$$
D \equiv \{ u \in C \vert~~ \vert u\vert < 1 \}.\eqno(2.8).
$$

The manifold
$G$
can be transformed into a Lie group relative to the following
composition law:
$$
(x,u)\cdot (y,v) \equiv \left( x+y+{1\over 2i} ln
{1+e^{-2iy}u\bar{v} \over 1+e^{2iy}v\bar{u}}, {u+e^{2iy}v \over
e^{2iy}+u\bar{v}} \right).\eqno(2.9)
$$

An explanation is needed. Let us denote:
$$
C_{r} \equiv \{ z \in C \vert~~ \vert z\vert = r \}~~~(\forall r
\in R_{+}).\eqno(2.10)
$$

Then if
$z \in D$
we have
$
{1 + z \over 1 + \bar{z}} \in C_{1} -\{-1\}
$
so
$
{1 + z \over 1 + \bar{z}}
$
can be uniquely written as
$
e^{2it}
$
with
$
t \in \left( -\pi /2, \pi/2 \right)
$.
It is natural to put
$$
t = {1 \over 2i} ln {1 + z \over 1 + \bar{z}}
$$
and this explains (2.9). We note that the same convention can be
applied for
$z$
pure imaginary.

The group
$G$
is the universal covering group of
$SL(2,R)$.
Indeed one can verify that the map
$
\delta_{1}: G \rightarrow SL(2,R)
$
given by:
$$
\delta_{1}(x,u) \equiv {1 \over 2\sqrt{1 - \vert u\vert^{2}}}
\left( \matrix {e^{ix}(1+u) + e^{-ix}(1+\bar{u}) &
ie^{ix}(1-u) - ie^{-ix}(1-\bar{u}) \cr
-ie^{ix}(1+u) + ie^{-ix}(1+\bar{u})  &
ie^{ix}(1-u) + ie^{-ix}(1-\bar{u}) } \right)\eqno(2.11)
$$
is well defined and it is a homomorphism.

We note that:
$$
ker(\delta_{1}) = \{ (2\pi n,0) \vert n \in Z \} \cong Z.\eqno(2.12)
$$

Next we need the covering map of
$SL(2,R)$
onto
$L^{\uparrow}_{+}$.
To this purpose we introduce the
$2 \times 2$
matrices
$
\tau_{0},\tau_{1},\tau_{2}
$
as follows [3]:
$$
\tau_{0} = \left( \matrix{ 1 & 0 \cr 0 & 1 } \right),
\tau_{1} = \left( \matrix {1 & 0 \cr 0 & -1} \right),
\tau_{2} = \left( \matrix {0 & 1 \cr 1 & 0 } \right).\eqno(2.13)
$$

These matrices are a basis in the linear space
$H$
of
$2 \times 2$
real symmetric matrices, and we have the isomorphism:
$$
M \ni x \mapsto [x] \in H\eqno(2.14)
$$
where:
$$
[x] \equiv x^{0}\tau_{0} + x^{1}\tau_{1} + x^{2}\tau_{2} =
\left( \matrix {x^{0}+x^{1} & x^{2} \cr x^{2} &
x^{0}-x^{1}} \right).\eqno(2.15)
$$

Then we define for any
$
A \in SL(2,R), \delta_{2}(A) \in End(M)
$
by:
$$
[\delta_{2}(A) x] = A [x] A^{t}.\eqno(2.16)
$$

One proves that
$
\delta_{2}(A) \in L^{\uparrow}_{+}
$,
and
$\delta_{2}$
is a group homomorphism with:
$$
ker(\delta_{2}) = \{\pm 1\}.\eqno(2.17)
$$

Now it is clear that
$
\delta \equiv \delta_{2} \circ \delta_{1}
$
is a covering homomorphism
$
\delta: G \rightarrow L^{\uparrow}_{+}
$
with:
$$
ker(\delta) = \{ (\pi n,0) \vert n \in Z \} \cong Z.\eqno(2.18)
$$

Because
$G$
is a simply connected Lie group it follows that it can be taken
as the universal covering group of
$L^{\uparrow}_{+}$.

It is clear that the universal covering group of
$P^{\uparrow}_{+}$
can be taken as the inhomogeneous group
$$
in(G) \equiv G \times_{t} M\eqno(2.19)
$$
where the homomorphism
$t: G \rightarrow Aut(M)$
is:
$$
t_{x,u}(a) \equiv \delta (x,u) a.\eqno(2.20)
$$

2.3 We want to classify the unitary irreducible representations
of the semi-direct product
$in(G)$.
To fix the notations we remind the reader the content of Mackey
theorem. Let
$H \times_{t} A$
be a semi-direct product of the locally compact groups
$H$
and
$A$
which verify the second axiom of countability. Suppose that
$A$ is Abelian. Here:
$
t: H \rightarrow Aut(A)
$
is a group homomorphism. To classify all the unitary irreducible
representations of
$H \times_{t} A$
one goes through the following steps [1].

(a) One considers the dual
$\hat{A}$
of
$A$
and the action of
$H$
on it given by:
$$
(h\cdot \omega)(a) \equiv \omega(t_{h^{-1}}(a)).\eqno(2.21)
$$

(b) One computes all the $H$-orbits in
$\hat{A}$.
We suppose there exists a Borel cross section
$\Sigma \subset \hat{A}$
intersecting once every $H$-orbit.

(c) For
$\forall \omega \in Z$
one computes the ``little group'':
$$
H_{\omega} \equiv \{ h \in H \vert h\cdot\omega = \omega \}.\eqno(2.22)
$$

(d) We suppose that we know the complete list of unitary
irreducible representations of
$H_{\omega}, \forall \omega \in Z$.

(e) Let
$O \subset \hat{A}$
be a $H$-orbit in
$\hat{A}, \omega_{0} \equiv O \cap Z$
and
$\pi$
a unitary irreducible representation of
$H_{\omega}$
acting in the (complex) Hilbert space
$K$.
As it is well known [1] one can associate to every
$\pi$
an one
$(H,O,K)$-cocyle
$\phi^{\pi}$
i.e. a Borel map
$
\phi^{\pi}: G \times O \rightarrow U(K)
$
(here $U(K)$ is the group of unitary operators in $K$) such that
a.e. in $G \times O$
$$
\phi^{\pi}(h_{1},h_{2}\cdot\omega) \phi^{\pi}(h_{2},\omega) =
\phi^{\pi}(h_{1}h_{2},\omega)\eqno(2.23)
$$
and
$\forall h \in H_{\omega_{0}}$
$$
\pi(h) = \phi^{\pi}(h,\omega_{0}).\eqno(2.24)
$$

A convenient way to construct
$\phi^{\pi}$
is as follows. Let
$c: O \rightarrow H$
be a Borel section i.e. a Borel map such that
$\forall \omega \in O$,
$$
c(\omega)\cdot\omega_{0} = \omega.\eqno(2.25)
$$

Then we can take:
$$
\phi^{\pi}(h,\omega) = \pi(c(h\cdot\omega)^{-1} h c(\omega)).\eqno(2.26)
$$

(f) For every $H$-orbit $O$ and every unitary irreducible
representation $\pi$ of
$H_{\omega_{0}}$
in $K$ we consider the Hilbert space
$H \equiv L^{2}(O,d\alpha,K)$
(where $\alpha$ is a $H$-quasi-invariant measure on $O$) and
define:
$
W^{(O,\pi)}_{h,a}: H \rightarrow H
$
as follows:
$$
(W^{(O,\pi)}_{h,a}f(\omega) = \omega(a) (r_{h}(h^{-1}\cdot\omega))^{1/2}
\phi^{\pi}(h,h^{-1}\cdot\omega) f(h^{-1}\cdot\omega)\eqno(2.27)
$$
(where
$r_{h}(\cdot)$
is a version of the Radon-Nycodym derivative
${d\alpha^{h^{-1}} \over d\alpha}$).

Then $W$ is a unitary irreducible representation of
$H \times_{t} A$.

Mackey theorem asserts that if the orbit structure is smooth (see
[1]) then every unitary irreducible representation of
$H \times_{t} A$
is unitary equivalent to a representation of the form
$W^{(O,\pi)}$
and moreover to distinct couples
$(O,\pi) \not= (O',\pi')$
correspond representations
$W^{(O,\pi)}$
and
$W^{(O,\pi)}$
which are not unitary equivalent.

In our case
$H = G = \widetilde{L^{\uparrow}_{+}}$
and $A = M$.

2.4 Steps (a) and (b) are very similar to the usual case of 1+3
dimensions. Namely we can identify
$\hat{M} \cong M$
as follows:
$$
M \ni p \mapsto \chi_{p} \in \hat{M}\eqno(2.28)
$$
where:
$$
\chi_{p}(a) \equiv e^{i\{a,p\}}.\eqno(2.29)
$$

Then the action of
$G$
on
$\hat{M}$
is the usual one:
$$
(x,u)\cdot p \equiv \delta (x,u) p\eqno(2.30)
$$
and the $G$-orbits in
$\hat{M}$
are:

(A)
$$
X_{m}^{\eta} \equiv\{ p \in \hat{M} \vert \Vert p\Vert^{2} =
m^{2}, sign(p^{0}) = \eta \}\eqno(2.31)
$$
(for
$m \in R_{+} \cup \{0\}, \eta = \pm$)

(B)
$$
Y_{m} \equiv \{ p \in \hat{M} \vert \Vert p\Vert^{2} = - m^{2}\}
\eqno(2.32)
$$
(for
$m \in R_{+}$)

(C)
$$
X_{00} \equiv \{0\}.\eqno(2.33)
$$

Here:
$$
\Vert p\Vert^{2} \equiv \{p,p\}.\eqno(2.34)
$$

We take as usual as the set of representative points:
$$
\Sigma \equiv \left( \cup_{m \in R_{+}} \{\pm me_{0}\} \right) \cup
\{\pm e_{+}\} \cup \left( \cup_{m \in R_{+}} me_{2}\right) \cup
\{0\}.\eqno(2.35)
$$

Here:
$$
e_{\pm} \equiv e_{0} \pm e_{1}.\eqno(2.36)
$$

2.5 The computation of the little groups
$H_{\omega}$
for
$\omega \in Z$
is elementary and we provide only the final results. We have;

(A)
$$
G_{\eta me_{0}} = \{ (x,0) \vert x \in R \} \cong R\eqno(2.37)
$$
(for $m \in R_{+}$) and:
$$
G_{\eta e_{+}} = \left\{ \left( {1 \over 2i} ln {1-ib \over 1+ib}
+n\pi, {ib \over 1-ib} \right) \vert n \in Z, b \in R\right\}.\eqno(2.38)
$$

Here the expresson
$ln {1-ib \over 1+ib}$
has been defined at 2.2. We note that we have:
$$
G_{\eta e_{+}} \cong Z \times R\eqno(2.39)
$$
where the isomorphism is:
$$
\left( {1 \over 2i} ln {1-ib \over 1+ib}
+n\pi, {ib \over 1-ib} \right) \leftrightarrow (n,b).\eqno(2.40)
$$

(B)
$$
G_{me_{2}} = \left\{ \left( \pi n,{a-1\over a+1}\right) \vert n
\in Z, a \in R_{+} \right\}.\eqno(2.41)
$$

We note that we have:
$$
G_{me_{2}} \cong Z \times R_{+}
$$
where
$R_{+}$
is considered as a multiplicative group and the isomorphism is:
$$
\left( \pi n,{a-1\over a+1}\right) \leftrightarrow (n,a).\eqno(2.42)
$$

(C) $G_{0} = G$.

2.6 The list of all the unitary irreducible representations for
the little groups above is very easy to determine.

(A) For
$G_{\eta me_{0}}$
these representations are indexed by a number
$s \in R$.
They are of the form:
$$
\pi^{s}(x,0) \equiv e^{isx}.\eqno(2.43)
$$

For
$G_{\eta e_{+}}$
one uses the group isomorphism (2.39), (2.40) and gets a list of
representations indexed by a couple
$(s,t)$
where
$s \in R(mod~ 2)$
and
$t \in R$
of the following form:
$$
\pi^{s,t}\left( {1 \over 2i} ln {1-ib \over 1+ib}
+n\pi, {ib \over 1-ib} \right) \equiv e^{\pi isn} e^{itb}.\eqno(2.44)
$$

(B) For
$G_{me_{2}}$
we get again representations indexed by a couple
$(s,t)$
where
$s \in R(mod~~2)$
and
$t \in R$
of the form:
$$
\pi^{',s,t}\left( \pi n,{a-1\over a+1}\right) \equiv
e^{\pi isn} a^{t}.\eqno(2.45)
$$

(C) The list of all unitary irreducible representations of
$G$
will be given in Section 3.

2.7 If we want explicit formulae for the corresponding
representations of the inhomogeneous group
$in(G)$,
it is necesery to determine the cocycles associated to the
representations of the little group identified above. In cases
(A) and (C) very simple formulae are available for a suitable
choices of the Borel section $c$.

(A) For the orbit
$X_{m}^{\eta}~~~(m \in R_{+}, \eta = \pm)$
one can show that a very simple Borel section
$c:X_{m}^{\eta} \rightarrow G$
is:
$$
c(p) = \left(0,{<p>\over p^{0}+\eta m}\right)~~~<p> \equiv
p^{1}+ip^{2}.\eqno(2.46)
$$

The corresponding cocycle
$\phi^{s}$
is:
$$
\phi^{s}((x,u),p) \equiv e^{isx} \left[ {p^{0}+\eta
m+ue^{-2ix}\bar{<p>} \over p^{0}+\eta
m+\bar{u}e^{2ix}<p>}\right]^{s/2}.\eqno(2.47)
$$

Here we interpret the expressions of the type
$\left( {a+z \over a+\bar{z}}\right)^{s/2}$
for
$a \in R_{+}$
and
$z \in C$
such that
$\vert z\vert < a$
as follows. We note that we have
${a+z \over a+\bar{z}} \not= -1$
so we can uniquely write
$ {a+z \over a+\bar{z}} = e^{2it}$
with
$t \in (-\pi/2,\pi/2)$.
Then we put
$\left( {a+z \over a+\bar{z}}\right)^{s/2} \equiv e^{ist}$.

We note that (2.47) is much more simpler than the corresponding
expression for the Wigner rotation obtained in [6] for the group
$P^{\uparrow}_{+}$. This is another benifit of working with the
covering group of
$P^{\uparrow}_{+}$.

For the orbit
$X_{0}^{\eta}$
a convenient cross section
$c:X_{0}^{\eta} \rightarrow G$
is:
$$
c(p) = \left(x_{0}(p),{p^{0}-1 \over p^{0}+1}\right)\eqno(2.48)
$$
where
$x_{0}(p) \in \left(-\pi/2,\pi/2\right)$
is determined by:
$$
x_{0}(p) \equiv Arg \left( \sqrt{{p^{0}+p^{1}} \over 2p^{0}} +
i \theta(p^{2}) \sqrt{{p^{0}-p^{1}} \over 2p^{0}}\right).\eqno(2.49)
$$

After some computations one can determine the corresponding
cocycle, namely:
$$
\phi^{'s,t}((x,u),p) = e^{isx} \left[ {p^{0}+
ue^{-2ix}\bar{<p>} \over p^{0}+
\bar{u}e^{2ix}<p>}\right]^{s/2}\times
$$
$$
exp\left\{it\eta{Im(ue^{-2ix}\bar{<p>})
\over p^{0}[(1+\vert u\vert^{2}) p^{0}+
2Re(ue^{-2ix}\bar{<p>})}\right\}.\eqno(2.50)
$$

(B) A determination of a Borel section
$c:Y_{m} \rightarrow G$
is still possible but the expression is very complicated so we
will not try to give explicit formulae in this case for the
corresponding cocycle
$\phi^{'s,t}$.

(C) It is clear that if
$\pi$
is a unitary irreducible representation of
$G$,
then we have:
$$
\phi^{\pi}((x,u),0) = \pi(x,u).\eqno(2.51)
$$

2.8 Applying Mackey theorem we end up with the following result:

{\bf Theorem 1:} Every unitary irreducible representation of
$in(G)$
is unitary equivalent to one of the following type:

(a) $W^{m,\eta,s}~~~(m \in R_{+}, \eta = \pm, s \in R)$
acting in
$L^{2}(X_{m}^{\eta},d\alpha^{\eta}_{m})$
as follows:
$$
\left( W^{m,\eta,s}_{x,u,a}f\right)(p) = e^{i\{a,p\}} e^{isx}
\left[ {p^{0}+\eta
m-\bar{u}e^{2ix}<p> \over p^{0}+\eta
m-ue^{-2ix}\bar{<p>}}\right]^{s/2} f((x,u)^{-1}\cdot p).\eqno(2.52)
$$

(b) $W^{\eta,s,t}~~~(\eta = \pm, s \in R(mod~~2), t \in R)$
acting in
$L^{2}(X_{0}^{\eta},d\alpha^{\eta}_{0})$
as follows:
$$
\left( W^{\eta,s,t}_{x,u,a}f\right)(p) = e^{i\{a,p\}} e^{isx}
\left[ {p^{0}
-\bar{u}e^{2ix}<p> \over p^{0}
-ue^{-2ix}\bar{<p>}}\right]^{s/2}\times
$$
$$
exp\left\{i\eta t{Im(ue^{-2ix}\bar{<p>}) \over
p^{0}[(1+\vert u\vert^{2}) p^{0}-
2Re(ue^{-2ix}\bar{<p>})}\right\}
f((x,u)^{-1}\cdot p).\eqno(2.53)
$$

(c) $W^{'m,s,t}~~~(m \in R_{+}, s \in R(mod~~2), t \in R)$
acting in
$L^{2}(Y_{m},\beta_{m},K)$
according to the formula:
$$
\left( W^{'m,s,t}_{x,u,a}f\right)(p) = e^{i\{a,p\}}
\phi^{'s,t}((x,u),(x,u)^{-1}\cdot p) f((x,u)^{-1}\cdot p).\eqno(2.54)
$$

Here
$\phi^{'s,t}$
is a cocycle corresponding to the representation
$\pi^{'s,t}$.

(d) $W^{\pi}$
acting in the Hilbert space
$K(\pi)$
of the unitary irreducible representation
$\pi$
of
$G$
according to:
$$
W^{\pi}_{x,u,a} = \pi(x,u).\eqno(2.55)
$$

(The measures
$\alpha^{\eta}_{m}~~~(m \in R_{+} \cup {0}), \eta = \pm)$
and
$\beta_{m}~~~(m \in R_{+} \cup \{0\}$
are defined as in [1].)

Two different representations in the list above are not unitary
equivalent.

2.9 {\bf Remarks}

1) From (2.52) and (2.53) we have a sort of Jacob-Wick limit
$m \rightarrow 0$:
$$
\lim_{m \rightarrow 0} W^{m,\eta,s} = W^{\eta,s,0}
$$

2) The infinitesimal generators are defined on a suitable G\aa rding
domain as follows:
$$
(P^{0}f)(p) \equiv -i{d\over dt}
(W_{0,0,te_{0}}f)(p)\vert_{t=0}.\eqno(2.56)
$$
$$
(P^{i}f)(p) \equiv i{d\over dt}
(W_{0,0,te_{i}}f)(p)\vert_{t=0}.\eqno(2.57)
$$
$$
(Jf)(p) \equiv i{d\over dt}
(W_{t/2,0,0}f)(p)\vert_{t=0}.\eqno(2.58)
$$
$$
(K^{1}f)(p) \equiv i{d\over dt}
(W_{0,t/2,0}f)(p)\vert_{t=0}.\eqno(2.59)
$$
$$
(K^{2}f)(p) \equiv i{d\over dt}
(W_{0,it/2,0}f)(p)\vert_{t=0}.\eqno(2.60)
$$

One can compute  explicitely these expressions for cases (a) and
(b) in the theorem above. It is convenient to identify functions
defined on
$X_{m}^{\eta}$
with functions defined on
$R^{2}$
taking into account the one-to-one correspondence:
$$
R^{2} \ni {\bf p} \leftrightarrow \tau({\bf p}) \in X_{m}^{\eta}\eqno(2.61)
$$
where:
$$
\tau({\bf p}) \equiv(E({\bf p}),{\bf p})\eqno(2.62)
$$
$$
E({\bf p}) \equiv \sqrt{{\bf p}^{2} + m^{2}}.\eqno(2.63)
$$

We get in both cases:
$$
P^{\mu} = \tau^{\mu}({\bf p}).\eqno(2.64)
$$
and
$$
J = i\left( p^{1} {\partial \over \partial p^{2}} -
p^{2} {\partial \over \partial p^{1}}\right) - s/2\eqno(2.65)
$$
For
$K^{i}$
one gets the following formulae:

-for $W^{m,\eta,s}$:
$$
K^{i} = i E({\bf p}) {\partial \over \partial p^{i}} +
s\varepsilon_{ij} {p^{j} \over 2[E({\bf p})+\eta m]}.\eqno(2.66)
$$

-for $W^{\eta,s,t}$:
$$
K^{i} = i E({\bf p}) {\partial \over \partial p^{i}} +
s\varepsilon_{ij} {p^{j} \over 2 E({\bf p})}+
{\eta t\over 2} \varepsilon_{ij} {p^{j}\over E({\bf p})^{2}}
.\eqno(2.67)
$$

These formulae should be compared with the similar ones obtained
in [6].

3) Let us note that by restriction to elements of the type
$(x,0,{\bf a}) \in G$
(i.e. to the universal covering group of the Euclidean group in
1+2 dimensions) we get in cases (a), (b) and (c) of the theorem
above a representation
$V$
acting in
$L^{2}(R^{2},d{\bf p})$
as follows:
$$
(V_{x,{\bf a}}f)({\bf p}) = e^{-i{\bf a}\cdot {\bf p}} e^{isx}
f(R(x)^{-1}{\bf p}).\eqno(2.68)
$$
where:
$$
R(x) \equiv \left( \matrix {cos(x) & -sin(x) \cr sin(x) & cos(x)}
\right).\eqno(2.69)
$$

Indeed, for cases (a) and (b) this is immediate.

In case (c) one uses the following argument. First one has the
direct integral decomposition:
$$
V = \int_{\mu \geq m} V^{\mu} \mu d\mu\eqno(2.70)
$$
where
$V^{\mu}$
acts in
$L^{2}(C_{\mu},d\varphi)$
as follows:
$$
(V^{\mu}_{x,{\bf a}}f)({\bf p}) = e^{-i{\bf a}\cdot {\bf p}}
\phi^{'s,t}((x,0),(x,0)^{-1}\cdot \tau({\bf p}))
f(R(x)^{-1}{\bf p}).\eqno(2.71)
$$

But the cocycle
$$
R \times C_{\mu} \ni x,{\bf p} \mapsto
\phi^{'s,t}((x,0),\tau({\bf p}) \in C\eqno(2.72)
$$
corresponds to a transitive action of the group
$R$
on the Borel space
$C_{\mu}$
so it is determined by its restriction to the stability subgroup
of, say,
$\mu e_{2}$
i.e.
$\{ \pi n \vert n \in Z \}$.
According to 2.6, this representation is
$\pi^{'s,0}$
(see (2.45)). Now it easily follows that the cocycle (2.72) is
equivalent to the cocycle
$\phi^{'s}$
given by:
$$
\phi^{'s}(x,{\bf p}) = e^{isx}.\eqno(2.73)
$$

So the representation
$V^{\mu}$
is equivalent to the representation (denoted also by
$V^{\mu}$),
acting in the same Hilbert space as follows:
$$
(V^{\mu}_{x,{\bf a}}f)({\bf p}) = e^{-i{\bf a}\cdot {\bf p}}
e^{isx} f(R(x)^{-1}{\bf p}).\eqno(2.74)
$$

{}From (2.70) and (2.74) it follows that (2.68) is valid for the
case (c) of the theorem above.

Now (2.68) shows like in [1] that the corresponding systems are
localizable on
$R^{2}$. Indeed,performing a Fourier transform
$F: L^{2}(R^{2},d{\bf p}) \rightarrow L^{2}(R^{2},d{\bf x})$
one brings (2.68) to the following form: $V$ is acting in
$L^{2}(R^{2},d{\bf x})$
according to:
$$
(V_{x,{\bf a}} f)({\bf x}) = e^{isx} f(R(x)^{-1} ({\bf x} - {\bf
a})).\eqno(2.75)
$$

Then, the corresponding projector-valued measure in
$L^{2}(R^{2},d{\bf x})$
is as usual
$\beta(R^{2}) \ni \Delta \mapsto \chi_{\Delta} \in
P(L^{2}(R^{2},d{\bf x}))$.

Also one can prove like in [1] that the systems corresponding to
case (d) in the theorem above are not localizable on
$R^{2}$.

4) One may wonder what is the ``spin'' of an elementary system
of the type (a), (b) or (c). For this one must first give a
cannonical definition of this notion. It is interesting to note
that one hardly finds a precise mathematical definition of this
kind stated in the litterature, although everybody has in mind
the following picture [1].

Suppose one has a certain {\it configuration space} i.e. a Borel
space $Q$ with a Borel action of some orthogonal group
$O(Q)$
on $Q$. Next, suppose that the Hilbert space of a certain
physical system is of the form
$H = L^{2}(Q,\alpha) \otimes K$
where
$\alpha$
is a quasi-invariant measure on $Q$ with respect to the action
of the group
$O(Q)$
and $K$ is a given Hilbert space.
Finally, suppose that our system is rotationally covariant i.e.
one has in $H$ a unitary representation of the universal
covering group
$\widetilde{O(Q)}$
of
$O(Q)$.
Moreover, this representation is supposed to be of the following
type:
$V = U \otimes W$
where
$$
(U_{g} f)(q) = r_{g}(q)^{1/2} f(\delta (g)^{-1} \cdot q).\eqno(2.76)
$$

Here
$r_{g}(\cdot)$
is a version of the Radon-Nycodym derivative
${d\alpha^{g^{-1}} \over d\alpha}$
and
$\delta: \widetilde{O(Q)} \rightarrow O(Q)$
is the covering homomorphysm.

Then we are entitled to say that the representation $U$ gives the
{\it orbital kinetic momentum} of the system and the
representation $W$ gives the {\it spin} of the system. In other
words we can identify the infinitesimal generators of $U$ and
$W$ with the orbital kinetic momentum and respectively with the
spin of the system.

Now we come to our specific situation of the representations
(a), (b) and (c) above. It is clear that one must make in the
general framework above the following particularizations:
$Q = R^{2}$,
$O(Q) = SO(2)$,
$\widetilde{O(Q)} \cong R$,
and
$K = C$.
The action of
$SO(2)$
on
$R^{2}$
is the usual one and the representation $W$ is simply:
$$
W_{x} = e^{isx}.\eqno(2.77)
$$

So we can conclude that in the cases (a), (b) and (c), the
spin of the system is $s$.

For a different point of view for the massless case (b), see
however [10].

2.10 One may wonder now what happens for the proper
orthochronous Galilei group
$G^{\uparrow}_{+}$
in 1+2 dimensions. In this case the analysis is much simpler
because the universal covering group for
$G^{\uparrow}_{+}$
has a much more simpler expression, namely is as a manifold:
$ \widetilde{G^{\uparrow}_{+}} = R \times R^{2} \times R \times R^{2}$
with the composition law:
$$
(x,{\bf v},\eta,{\bf a}) \cdot (x',{\bf v'},\eta',{\bf a'}) =
(x+x',{\bf v}+R(x){\bf v'},\eta+\eta',{\bf a}+R(x){\bf
a'}+\eta'{\bf v}),\eqno(2.78)
$$
where
$R(x)$
has been defined at (2.69). The covering homomorphism
$\delta:\widetilde{G^{\uparrow}_{+}} \rightarrow G^{\uparrow}_{+}$
is:
$$
\delta(x,{\bf v},\eta,{\bf a}) = (R(x),{\bf v},\eta,{\bf a}).\eqno(2.79)
$$

Now the analysis is straightforward and can be obtained by
appropriate modifications of the similar analysis from [1]
dedicated to the 1+3 dimensional case. We give only the final result:

{\bf Theorem 2:} Every unitary irreducible representation of
$\widetilde{G^{\uparrow}_{+}}$
is equivalent to one of the folowing type:

(a) $V^{m,s}~~~(m \in R^{*}, s \in R)$
acting in
$L^{2}(R^{2},d{\bf p})$
as follows:
$$
(V^{m,s}_{x,{\bf v},\eta,{\bf a}}f)({\bf p}) = exp\left\{i\left( sx+
{\bf a}\cdot{\bf p}+{\eta{\bf p}^{2} \over 2m}+{m \over 2}
{\bf a}\cdot{\bf v}\right)\right\} f(R(x)^{-1}({\bf p}+
m{\bf v})).\eqno(2.80)
$$

(b) $L^{r,s,t}~~~(r \in R_{+}, s \in R(mod~~1), t \in R)$
acting in
$L^{2}(R \times C_{r}, dp_{0} \times d\varphi)$
as follows:
$$
(L^{r,s,t}_{x,{\bf v},\eta,{\bf a}}f)(p_{0},{\bf p}) = exp\left\{i\left( sx+
{\bf a}\cdot{\bf p}+\eta p_{0} +{t [{\bf p},{\bf v}]\over r}
\right)\right\} f(p_{0}+{\bf v}\cdot{\bf p},R(x)^{-1} {\bf p})
.\eqno(2.81)
$$

Here
$\forall {\bf a}, {\bf b} \in R^{2}$
we have defined the antisymmetric bilinear form
$[\cdot,\cdot]$
by:
$$
[{\bf a},{\bf b}] e_{1} \wedge e_{2} \equiv {\bf a} \wedge {\bf b}
.\eqno(2.82)
$$

(c) $L^{p_{0},\rho,s}~~~(p_{0} \in R, \rho \in R_{+}, s \in
R(mod~~1))$
acting in
$L^{2}(C_{\rho},d\varphi)$
as follows:
$$
(L^{p_{0},\rho,s}_{x,{\bf v},\eta,{\bf a}}f)(\omega) = exp\left\{i\left( sx+
\eta p_{0}+\omega\cdot{\bf v}\right)\right\} f(R(x)^{-1} \omega)
.\eqno(2.83)
$$

(d) $L^{p_{0},s}~~~(p_{0} \in R, s \in R)$
acting in $C$ as follows:
$$
L^{p_{0},s}_{x,{\bf v},\eta,{\bf a}} = exp\{i(\eta p_{0}+sx)\}.\eqno(2.84)
$$

{\bf Remarks:}

1) The representation
$L^{p_{0},s}$
induces a trivial projective representation of
$G^{\uparrow}_{+}$
so it can be discarded.

2) As in [1] one can show that only
$V^{m,s}$
is localizable on
$R^{2}$.
\vskip1 truecm
{\bf 3. The Unitary Irreducible Representation of the }

{\bf Universal Covering Group of} $SL(2,R)$

3.1 The unitary irreducible representations of
$SL(2,R) \cong SU(1,1)$
are rather well studied in the litterature [8], [11], [12]. The
corresponding problem for the universal covering group is
treated in [4], [5]. However, as pointed out in the Introduction,
Ref. [4] does not use the explicit construction for the covering
group and the classification of the unitary irreducible
representations is done in an implicit way without explicit
realizations. (Also [5] is focused on
$SU(1,1)$
and the formulae are rather cumbersome).
The same can be said about the analysis of the
discrete series appearing in e.g. [6], [7] where only the
infinitesimal generators appear.

We find it profitable for the reader to provide here all the
relevant formulae for the universal covering group of
$SL(2,R)$. Of course we will reproduce
(in the next Subsection) a large part of the results of [4].

3.2 We identify
$Lie(G) = Lie(SL(2,R))$
with the three dimensional space of
$2 \times 2$
real traceless matrices. A basis in this space is
$\{l_{0}, l_{1}, l_{2}\}$
where (see (2.13)):
$$
l_{0} \equiv -{1 \over 2} \tau_{3},~~~l_{1} \equiv {1 \over 2}
\tau_{1}, ~~~l_{2} \equiv {1 \over 2} \tau_{2}.\eqno(3.1)
$$

We have the commutations relations:
$$
[l_{0},l_{1}] = l_{2},~~~[l_{0},l_{2}] = -l_{1},~~~[l_{1},l_{2}]
= -l_{0}.\eqno(3.2)
$$

Let $T$ be a unitary irreducible representation of the group $G$
in the Hilbert space $H$. For any
$l \in Lie(G)$
we denote by
$H_{l}$
the self-adjoint operator in $H$ determined (according to
Stone-von Neumann theorem) by:
$$
e^{-itH_{l}} = T_{exp(tl)}~~~(\forall t \in R).\eqno(3.3)
$$

Let us denote by
$D_{l}$
the domain of self-adjointness of
$H_{l}$.
Then
$B \equiv D_{l_{0}} \cap D_{l_{1}} \cap D_{l_{2}}$
is a G\aa rding domain for $T$, and we have in $B$:
$$
[H_{0},H_{1}] = i H_{2},~~~[H_{0},H_{2}] = -
iH_{1},~~~[H_{1},H_{2}] = -iH_{0}.\eqno(3.4)
$$

It is well known that the representation $T$ is uniquely
determined by the infinitesimal generators
$H_{i} (i = 0,1,2)$.
The usual way to proceed for a non-compact group is to search
for a maximal compact subgroup for which the corresponding
infinitesimal generators will have a pure point spectrum. In the
case of our group $G$ the maximal compact subgroup is trivially
formed by the neutral element, so appearently we cannot procced
further. However a trick of [4] shows that the spectrum of
$H_{0}$ is pure point.

Next one defines on $B$ the operator
$$
Q' \equiv (H_{1})^{2} + (H_{2})^{2} - (H_{0})^{2}\eqno(3.5)
$$
and by $Q$ the unique self-adjoint extension to $H$. Like in [7]
one proves that:
$$
Q = q~I~~~(q \in R).\eqno(3.6)
$$

If we denote by
$D \subset H$
the linear subspace of finite linear combinations of
eigenvectors of
$H_{0}$,
then one can show that $D$ is a G\aa rding domain for $T$ [8]. So
it will be sufficient to determine the action of
$H_{i}~~~(i = 0,1,2)$
on $D$.

Let us denote:
$$
H_{\epsilon} \equiv H_{1} + i\epsilon H_{2}~~(\epsilon = \pm).\eqno(3.7)
$$

Then the final result of the infinitesimal analysis is the
following. The commutation relations (3.4) are compatible with
the folowing three cases:

I. There exists
$f \in D$
such that
$H_{\epsilon}^{j} f \not= 0~~~(\forall \epsilon = \pm, \forall j
\in N)$.
In this case one can find
$\tau \in [0,1)$
and an orthonormal base in $H$
$\{f_{\alpha}\}_{\alpha \in \tau + Z}$
such that:
$$
H_{0} f_{\alpha} = \alpha f_{\alpha}\eqno(3.8)
$$
$$
H_{\epsilon} f_{\alpha} = [q + \alpha (\alpha + \epsilon)]^{1/2}
f_{\alpha+\epsilon}.\eqno(3.9)
$$

Moreover one must have
$$
q > \tau (1 - \tau).
$$

II. There exists
$f \in D$
such that
$f \not= 0$
but
$H_{-} f = 0$.
Then one can find
$l \in R_{+}$
and an orthonormal base
$\{f_{\alpha}\}_{\alpha \in l + N}$
in $H$ such that (3.8) and (3.9) stay true for the appropriate
values of
$\alpha$
(we also take
$f_{l-1} \equiv 0$).

III. There exists
$f \in D$
such that
$f \not= 0$
but
$H_{+} f = 0$.
Then one can find
$l \in R_{+}$
and an orthonormal base
$\{f_{\alpha}\}_{\alpha \in -l - N}$
in $H$ such that (3.8) and (3.9) stay true for the appropriate
values of
$\alpha$
(we also take
$f_{-l+1} \equiv 0$).

3.3 Next one proves in [4] in a very implicit way that to every
case above one has indeed a unitary irreducible representation
of $G$.

We will prove this point by some very explicit constructions.
The basic result is:

{\bf Theorem 3:} Let
$H = L^{2}(2\pi,d\varphi)$
and
$s \in R(mod~~2), t \in R)$.
For any
$(x,u) \in G$
let us define the linear operator
$T_{x,u}:H \rightarrow H$
as follows:
$$
(T_{x,u}^{s,t} f)(\varphi) = e^{is\varphi} \left( {1 -
\bar{u}e^{i(\varphi-2x)} \over 1 - ue^{-i(\varphi-2x)}}\right)^{s/2}
\left( {\vert1 - ue^{-i(\phi-2x)}\vert \over \sqrt{1 - \vert u\vert^{2}}}
\right)^{it-1} \times
$$
$$
f\left( \varphi - 2x + {1 \over i} ln{1 -
ue^{-i(\varphi-2x)} \over 1 - \bar{u}e^{i(\varphi-2x)}}\right).\eqno(3.10)
$$

Here $f$ is extended to the whole real axis by periodicity with
period
$2\pi$.

Then $T$ is a unitary representation of $G$ in $H$ corresponding
to
$\tau = -{s \over 2} \in [0,1)$
and
$q = {1\over 4} (t^{2} +1) \geq {1\over4}$.

{\bf Proof:} The fact that
$T_{x,u}^{s,t}$
is unitary follows by a convenient change of variables in the
expression
$\Vert T_{x,u}^{s,t} f\Vert^{2}$
and taking into acount the periodicity of $f$. For the
representation property one can avoid a brute force computation
rewritting a little bit (3.10).

First, we consider $f$ as a function defined on the circle
$C_{1} = \{ (\zeta^{1})^{2}+(\zeta^{2})^{2} = 1 \}$
taking:
$$
\zeta^{1} = cos(\varphi),~~~\zeta^{2} = sin(\varphi).\eqno(3.11)
$$

Next one defines
$\forall \zeta \in C_{1},~~\tau(\zeta) \in X_{0}^{\uparrow}$
by:
$$
\tau(\zeta) \equiv (1,\zeta^{1},\zeta^{2}).\eqno(3.12)
$$

Finally, we define
$b:X_{0}^{\uparrow} \rightarrow C_{1}$
as follows:
$$
b(p) \equiv {p^{1} + i p^{2} \over p^{0}} = {<p> \over p^{0}}.\eqno(3.13)
$$

Now one must check that (3.10) is the same as:
$$
(T_{x,u}^{s,t} f)(\zeta) = \phi^{s}((x,u),(x,u)^{-1}\cdot
\tau(\zeta)) \vert<(x,u)\cdot\tau(\zeta)>\vert^{it-1}
f(b((x,u)^{-1}\cdot\tau(\zeta))b(\tau(\zeta))^{-1}).\eqno(3.14)
$$
where the cocyle
$\phi^{s}$
has been defined at (2.47). The new realization (3.14) makes the
representation preperty rather easy to establish.

It remains to match
$T^{s,t}$
with one of the cases I, II, III from the preceeding Subsection.
For this we need the infinitesimal generators of $T$; they are
rather easily computed:
$$
H_{0} = - \left( i{d \over d\varphi} + {s \over 2}\right)\eqno(3.15)
$$
$$
H_{\epsilon} = e^{i\epsilon\varphi} \left( i\epsilon H_{0} + {t
+ i\over 2}\right).\eqno(3.16)
$$

Let us take for any
$n \in Z$:
$$
g_{n}(\varphi) \equiv {1 \over \sqrt{2\pi}} e^{in\varphi}.\eqno(3.17)
$$

Then, according to Fourier theorem,
$\{g_{n}\}_{n \in Z}$
is an orthonormal basis in $H$. Moreover we immediately have:
$$
H_{0} g_{n} = \left(n - {s\over 2}\right) g_{n}\eqno(3.18)
$$
$$
H_{\epsilon} g_{n} = \left[ i\epsilon \left(n - {s\over
2}\right) + {t + i\over 2} \right] g_{n+\epsilon}\eqno(3.19)
$$
$$
Q = {1 \over 4} (t^{2} + 1) I.\eqno(3.20)
$$

Comparing with case I it is evident that one should take
$s = -2\tau$
(with
$\tau \in [0,1))$
and
$q = {1\over 4} (t^{2} + 1)$.
The condition
$q > \tau(1 -\tau)$
is always true. If we redefine
$g_{n} \rightarrow g_{n+\tau}$
we get from (3.18) and (3.19) a set of relations more familiar
with I, namely:
$$
H_{0} g_{\alpha} = \alpha g_{\alpha}\eqno(3.21)
$$
$$
H_{+} g_{\alpha} = \omega_{\alpha} \sqrt {q + \alpha(\alpha+1)}
g_{\alpha+1}\eqno(3.22)
$$
$$
H_{-} g_{\alpha} = {1\over\omega_{\alpha-1}} \sqrt {q + \alpha(\alpha-1)}
g_{\alpha-1}\eqno(3.23)
$$

Here
$$\omega_{\alpha}\equiv{i\alpha+ {t+i\over 2}\over\sqrt{q
+\alpha(\alpha+1)}} \eqno(3.24)
$$
is a complex number of modulus 1.

It is sufficient to take now
$\{\epsilon_{\alpha}\}_{\alpha \in \tau+Z}$,
such that
$\vert\epsilon_{\alpha}\vert = 1$
and
$\omega_{\alpha} = {\epsilon_{\alpha} \over \epsilon_{\alpha-1}}$
and to define the new orthonormal base
$\{f_{\alpha}\}_{\alpha \in \tau+Z}$
by
$f_{\alpha} \equiv \epsilon_{\alpha}^{-1} g_{\alpha}$;
then (3.21)-(3.23) go into (3.8), (3.9). Q.E.D.

We will also denote
$T^{s,t}$
by
$T^{\tau,q}$.
According to the standard terminology the representations
$T^{\tau,q}~~~(q \geq{1/4})$
constitute the {\it principal series}.

Let us remark that
$T^{\tau,q}$
induces a true representation of
$SL(2,R)$
{\it iff} $\tau = 0, 1/2$
and induces a true representationof
$L^{\uparrow}_{+}$
{\it iff} $\tau = 0$.

3.4 It is clear from the preceeding theorem that the case
$q < 1/4$
can be obtained making formally
$t \rightarrow it$.
Of course this modification will ruin the unitarity. The idea is
to modify appropriately the expression of the scalar product [4],
[12]. We define on the space
$V$
of smooth complex periodic functions with period
$2\pi$
the representation of $G$:
$$
(T_{x,u}^{'s,t} f)(\varphi) = e^{is\varphi} \left( {1 -
\bar{u}e^{i(\varphi-2x)} \over 1 - ue^{-i(\varphi-2x)}}\right)^{s/2}
\left( {\vert1 - ue^{-i(\phi-2x)}\vert \over \sqrt{1 - \vert u\vert^{2}}}
\right)^{-t-1}\times
$$
$$
f\left( \varphi - 2x + {1 \over i} ln{1 -
ue^{-i(\varphi-2x)} \over 1 - \bar{u}e^{i(\varphi-2x)}}\right).\eqno(3.25)
$$
where
$s \in R(mod~~2)$
and $t \in R_{+}$.

Then, we define on $V$ the sesquilinear form
$<\cdot,\cdot>$
by:
$$
<f,g> \equiv \int_{0}^{2\pi} \int_{0}^{2\pi} d\varphi d\varphi'
L(\varphi - \varphi') \bar{f(\varphi)} g(\varphi')\eqno(3.26)
$$
where the kernel $L$ is:
$$
L(\varphi) \equiv 2^{t}\pi e^{i\tau \pi} B\left( \tau + {1+t\over 2},
-\tau + {1+t\over 2}\right) \vert sin(\varphi /2)
\vert^{t-1} e^{-i\tau \varphi}.\eqno(3.27)
$$

Like above we take:
$s = -2\tau$
(with
$\tau \in [0,1)$)
and
$q = {1\over 4} (1 - t^{2})$.
The condition
$q > \tau(\tau-1)$
gives
$t < \vert 1-2\tau\vert$.

One proves that
$<\cdot,\cdot>$
is non-degenerated as follows. We define for any
$n \in Z$
the functions
$g_{n} \in V$
by:
$$
g_{n}(\varphi) = \gamma_{n} e^{in\varphi}\eqno(3.28)
$$
where:
$$
\gamma_{n} \equiv \left[ { \Gamma\left( n+\theta (n)\tau+{1+t\over
2}\right) \Gamma\left(\theta (n)\tau +{1-t\over 2}\right) \over
\Gamma\left( n+\theta (n)\tau+{1-t\over
2}\right) \Gamma\left(\theta (n)\tau +{1+t\over 2}\right)}
\right]^{1/2} .\eqno(3.29)
$$

Then using [13] ( \$ 3.63) one can show that:
$$
L(\varphi) = \sum_{n \in Z} {1 \over \gamma_{n}^{2}} e^{in\varphi}.\eqno(3.30)
$$

It follows easily that
$\{g_{n}\}_{n \in Z}$
is an orthonormal system in $V$ and
$<\cdot,\cdot>$
is positively defined.
So we can obtain from $V$, by completion, a Hilbert space $H$
(in which
$\{g_{n}\}_{n \in Z}$
is an orthonormal basis) and extend
$T^{'s,t}$
by continuity to $H$. It is not very hard to prove that
$T^{'s,t}$
is unitary with respect to the scalar product (denoted also by
$<\cdot,\cdot>$)
of $H$.

{}From (3.18)-(3.20) with
$t \rightarrow it$
we obtain
($g_{n} \rightarrow g_{n+\tau}$):
$$
H_{0} g_{\alpha} = \alpha g_{\alpha}\eqno(3.31)
$$
$$
H_{\epsilon} g_{\alpha} = i\epsilon~~\eta_{\alpha} \sqrt {q + \alpha(\alpha+
\epsilon)} g_{\alpha+\epsilon}\eqno(3.32)
$$
where:
$$
\eta_{\alpha} \equiv \cases {1 &for $\alpha \in \tau+N$ \cr
sign\left( {1+t\over 2} - \tau\right) &for $\alpha \in \tau
- N^{*}$.}\eqno(3.33)
$$

Now let
$\{\epsilon_{\alpha}\}_{\alpha \in \tau+N}$
such that
$\vert\epsilon_{\alpha}\vert = 1$
and
${\epsilon_{\alpha +1}\over \epsilon_{\alpha}} = i\eta_{\alpha}$.
If we redefine
$f_{\alpha} = \epsilon_{\alpha}^{-1} g_{\alpha}$
then (3.31) and (3.32) give (3.8) and (3.9).

We will also denote
$T^{'s,t}$
by
$T^{\tau,q}$.
The representations
$T^{\tau,q}~~~(q < 1/4)$
constitute the so-called {\it complementary series}. In this
case we get true representations for
$SL(2,R)$
{\it and} for
$L^{\uparrow}_{+}$
{\it iff}
$\tau = 0$.

3.5 The cases II and III from 3.2 are settled by:

{\bf Theorem 4:} Let $F$ be the linear space of all analytic
functions on the disk
$D = \{ z \in C \vert~~\vert z\vert < 1 \},~~l \in R_{+}$
and
$\eta = \pm$.
For any
$(x,u) \in G$
we define
$D^{l,\eta}_{x,u}: F \rightarrow F$
as follows:
$$
(D^{l,+}_{x,u}f)(z) = e^{-2ilx} (1 + e^{-2ix}\bar{u} z)^{-2l} (1
- \vert u\vert^{2})^{l} f\left( {z+e^{2ix} u\over
e^{2ix}+\bar{u} z}\right)\eqno(3.34)
$$
and respectively
$$
(D^{l,-}_{x,u}f)(z) = e^{2ilx} (1 + e^{2ix}u z)^{-2l} (1
- \vert u\vert^{2})^{l} f\left( {z+e^{-2ix} \bar{u}\over
e^{-2ix}+uz}\right)\eqno(3.35)
$$

We also define on $F$ the Hermitean form
$<\cdot,\cdot>$
by:
$$
<f,g> \equiv {2l\over \pi} \int_{D} (1 - \vert z\vert^{2})^{2(l-1)}
\bar {f(z)} g(z) d\sigma\eqno(3.36)
$$
($d\sigma$
is the surface measure on $C$). Let $H$ be the Hilbert space
obtained from $F$ by completion with respect to
$<\cdot,\cdot>$,
and extend
$D^{l,\eta}$
to $H$ by continuity. Then $T$ is a unitary representation of
$G$ in $H$ corresponding to cases II and III for
$\eta = +$
and
$\eta = -$
respectively.

{\bf Remark} In (3.34) and (3.35) we interpret
$(1+u)^{s}$
for
$\vert u\vert < 1$
taking
$Arg(1+u) \in (-\pi,\pi)$.

{\bf Proof:} We first remark that
$D^{l,+}$
and
$D^{l,-}$
can be obtained one from the other by complex cojugation and the
substitution
$z \rightarrow \bar{z}$.

There are a number of fact which can be easily verified by
direct computation namely that
$D^{l,\eta}$
is well defined, leaves
$<\cdot,\cdot>$
invariant and have the repesentation property.

It remains to match these representations with the infinitesimal
analysis. We do this for
$D^{l,+}$.
We get immediately for the infinitesimal generators:
$$
H_{0} = z{d\over dz} + l.\eqno(3.37)
$$
$$
H_{+} = -2ilz -iz^{2} {d\over dz}\eqno(3.38)
$$
$$
H_{-} = i {d\over dz}.\eqno(3.39)
$$

If we define
$\forall n \in N$
the functions
$g_{n} \in F$ by:
$$
g_{n}(z) \equiv (-1)^{n} z^{n}\eqno(3.40)
$$
then we get:
$$
H_{0} g_{n} = (l+n) g_{n}\eqno(3.41)
$$
$$
H_{+} g_{n} = i(n+2l) g_{n+1}\eqno(3.42)
$$
$$
H_{-} g_{n} = -in g_{n-1}.\eqno(3.43)
$$

Now we consider the functions
$f_{n} \in F$
given by:
$$
f_{n} \equiv \gamma_{n} g_{n}\eqno(3.44)
$$
where
$$
\gamma_{n} \equiv \left[{ \Gamma (n+2l) \over \Gamma (2l) \Gamma
(n+1)} \right]^{1/2}.\eqno(3.45)
$$

It is easy to prove that
$\{f_{n}\}_{n \in N}$
is an orthonormal basis in $H$ and that (3.41)-(3.43) give
(3.8), (3.9) if we make
$f_{n} \rightarrow f_{l+n}$.

For
$D^{l,-}$
one obtains the infinitesimal generatores making
$H_{0} \rightarrow H_{0}, H_{\epsilon} \rightarrow
-(H_{\epsilon})^{*}$.
The analysis above stays true if we take
$\forall n \in -N$:
$$
g_{n}(z) \equiv (-1)^{n} z^{-n}\eqno(3.46)
$$
and in (3.45) make
$n \rightarrow -n$. Q.E.D.

The representations
$D^{l,\eta}$
constitute the so-called {\it discrete series}. As remarked in
the Introduction they can be obtained with the help of geometric
quantization [6]. Let us note that they induce true
representations for
$SL(2,R)$
{\it iff}
$l \in {N^{*} \over 2}$,
and one easily obtains the formulae of [10].
They induces true representations for
$L^{\uparrow}_{+}$
{\it iff}
$l \in N^{*}$.

3.6 From Subsections 2.2-2.5 we conclude that
$T^{\tau,q}~~~(\tau \in [0,1), q > \tau(1-\tau) )$
and
$D^{l,\eta}~~~(l \in R_{+}, \eta = \pm) $
are all the unitary irreducible representations of $G$, up to a
unitary equivalence.
\vskip 1truecm
{\bf 4. Conclusions}

We have given a complete list for all the projective unitary
irreducible representations of the Poincar\'e group in 1+2
dimensions, up to unitary equivalence. Except for the case of
tachyons we have been able to produce very explicit formulae
which can be of practical use in various applications.

It will be interesting to proceed further to field theory and
develop the appropriate generalizations for invariant wave
equations (see [6], [7]), Fock-Cook formalism for an arbitrary
statistics and the basic theorems of axiomatic field theory.
These subjects will be approached elsewere.
\vskip 1truecm
\vfil\eject

{\bf References}

\item{1.}
V. S. Varadarajan, ``Geometry of Quantum Theory'' (second edition),
Springer, 1985
\item{2.}
V. Bargman, ``On Unitary Ray Representations of Continous
Groups'', Ann. Math. 59 (1954) 1-46
\item{3.}
B Binegar, ``Relativistic Field Theory in Three Dimensions'', J.
Math. Phys. 23 (1982) 1511-1517
\item{4.}
L. Puk\'ansky, ``The Plancherel Formula for the Universal
Covering Group of $SL(2,R)$'', Math. Annalen 156 (1964) 96-143
\item{5.}
P. J. Sally, Jr., ``Analytic Continuation of the Irreducible
Unitary Representations of the Universal Covering Group of
$SL(2,R)$'',
Mem. Amer. Soc. 69 (1967) 1-94
\item{6.}
R. Jackiw, V. P, Nair, ``Relativistic Wave Equations for
Anyons'', Phys. Rev. D 43 (1991) 1933
\item{7.}
M. S. Plyushchay, ``Relativistic Particle with Torsion, Majorana
Equation and Fractional Spin'', Phys. Lett. B 262 (1991) 71-78
\item{8.}
V. Bargaman, ``Ireducible Unitary Representations of the Lorentz
Group'', Ann. Math. 48 (1947) 568-640
\item{9}
M. Postnikov, ``Le\c cons de G\'eom\'etrie. Groupes et Alg\`ebres de
Lie'', ed. Mir, Moscou, 1985
\item{10.}
S. Deser, R. Jackiw, ``Statistics without Spin'', Phys. Lett. B
263 (1991) 431-436
\item{11.}
S. Lang, ``$SL_{2}(R)$'' (scond edition), Springer, 1985
\item{12.}
I. M. Gelfand, M. I. Graev, N. Ya. Vilenkin, ``Generalized
Functions, vol. 5. Integral Geometry and Representation
Theory'', Academic Press, 1966
\item{13.}
I. S. Gradshteyn, I. M. Ryzik, ``Tables of Integrals, Series and
Products'', Academic Press, 1980
\bye